\documentclass[prl,twocolumn,amsmath,amssymb,groupedaddress,superscriptaddress]{revtex4}
\usepackage{graphics}
\usepackage{amsmath}
\usepackage[dvips]{graphicx}
\usepackage{float,graphicx}
\usepackage{subfigure,hyperref}

\def\>{\rangle}
\def\<{\langle}

\newcommand{\unit}{1\!\!1}

\def\be{\begin{eqnarray}}
\def\beqa{\begin{eqnarray}}

\def\ee{\end{eqnarray}}
\def\eeqa{\end{eqnarray}}

\def\mmax{\textmd{m}}

\def\ttot{\textmd{tot}}

\def\ttot{\textmd{tot}}

\def\Nsl{\textmd{NSL}}
\def\Sl{\textmd{SL}}
\def\Ttr{\textmd{Tr}}

\def\Hhc{\textmd{H.c.}}

\def\tth{\textmd{th}}

\def\be{\begin{equation}}
\def\ee{\end{equation}}
\def\bea{\begin{eqnarray}}
\def\eea{\end{eqnarray}}
\usepackage{bm}
\usepackage{graphicx}
\begin{document}
\bibliographystyle{unsrt}
\title[]{A weak-coupling master equation for arbitrary initial conditions}
\author{Jad C. Halimeh}
\affiliation{Department of Physics and Arnold Sommerfeld Center for Theoretical Physics, Ludwig-Maximilians-Universit{\"a}t M{\"u}nchen, Theresienstr. 37, 80333 Munich, Germany}
\author{In\'es de Vega}
\affiliation{Department of Physics and Arnold Sommerfeld Center for Theoretical Physics, Ludwig-Maximilians-Universit{\"a}t M{\"u}nchen, Theresienstr. 37, 80333 Munich, Germany}
\date{\today}
\begin{abstract}
The structure of the initial system-environment state is fundamental to determining the nature and characteristics of the evolution of such an open quantum system. The usual assumption is to consider that the initial system-environment state is separable. Here, we go beyond this simple case and derive the evolution equations, up to second order in a weak-coupling expansion, that describe the evolution of the reduced density matrix of the system for any arbitrary system-environment initial state. The structure of these equations allows us to determine the initial conditions for which a Lindblad form can be recovered once applying the Markov and secular approximations. 
\end{abstract}
\maketitle
The dynamical properties of the reduced density matrix of an open quantum system (OQS) are strongly dependent on the initial condition considered \cite{devega2015c}. In this regard, Pechukas  \cite{pechukas1994} was the first to point out that initially correlated states between the system and its environment might lead to dynamics that do not preserve the property of complete positivity. This idea was the subject of an intense debate with Alicki \cite{alicki1995}, who argued that all physically meaningful initial states should lead to complete positive dynamics. Since those early papers, this subject has been the basis of many works that aim at formally determining which types of initial state lead to certain desirable properties such as linearity or complete positivity \cite{modi2012a,rosario2010,jordan2004,carteret2008,rosario2008,shabani2009,brodutch2013,buscemi2014,erratumshabani}.
The existence of initial correlations may modify not only the system dynamics, but also the exchange of information between the system and the environment that exists during the decoherence process. In this regard, an initial flow of information from the environment to the system has been linked to the presence of such initial correlations \cite{laine2010} (see also \cite{breuer2015,devega2015c} and references therein). 
More recently, the occurrence of system-environment correlations generated during the evolution has been linked to non-Markovianity \cite{mazzola2012,smirne2013}. 

%
%
%

Many previous studies of system dynamics for an initially correlated state consider an exactly solvable model \cite{smirne2010,laine2010,lazarou2011,hua-tang2011}. Other proposals analyze the reduced dynamics of an OQS considering an initial thermal state for the full system and expanding it into a first term, which is separable, and a second term, which accounts for the correlations and is of first order in the coupling parameter between system and environment \cite{chaudhry2013,chen2016}. This allows one to consider standard projection operator techniques in combination with a weak-coupling expansion. A third approach is based on correlated projection operators \cite{esposito2003,budini2006,breuer2007,breuerbook,devega2015c}, where a fixed form for the relevant (in the sense that its trace over the environment gives rise to a good approximation of the reduced density operator of the system) part of the total density operator is assumed as an ansatz. In addition, the reduced hierarchical equations of motion, which are particularly adapted to exponentially decaying environment correlation functions, have also been extended to deal with correlated initial conditions \cite{tanimura2014}.

Here, we adopt a different approach that is based on considering a reduced propagator of the evolution operator in a Bargmann coherent state basis as defined in \cite{strunz2001,alonso2005,devega2006}, and on a weak-coupling approximation between system and environment. Based on this, we derive an evolution equation for the reduced density matrix that is valid for any arbitrary initial state and for any environment spectral density, and is also free from any ans\"atze. Moreover, the resulting equation is deterministic, \emph{i.e.} its practical usage does not rely on a stochastic sampling like the derivation in \cite{devega2006}. 
The analysis presented here can also be straightforwardly extended to multi-level or many body OQSs. 
In addition, considering this equation in the Markov and secular approximation limits, we show what types of initial states (do not) give rise to an equation that is of a Lindblad form, and, therefore, (do not) preserve complete positivity. 



Let us consider an OQS with Hamiltonian $H_S$, linearly interacting with a harmonic oscillator environment described by a Hamiltonian $H_B=\sum_\lambda \omega_\lambda a_\lambda^\dagger a_\lambda $, such that the total Hamiltonian is 
\begin{eqnarray}
H_\ttot&=&
H_S + H_B+ B^\dagger  L  + B L^\dagger,
\label{chapdos1}
\end{eqnarray}
with $B=\sum_\lambda g_\lambda a_\lambda$ an environment operator that depends on the ladder operators $a_\lambda,a_\lambda^\dagger$, and $L$ is an operator acting on the Hilbert space of the system. The $g_\lambda$'s are the coupling constants, which can be made real, and the $\omega_\lambda$'s are the frequencies of the environment harmonic oscillators. 
%
Let us consider initial states of the general form
\bea
\rho_\ttot(0)=\sum_{\gamma=1}^M \phi_s^\gamma(0)\otimes\phi^\gamma_B(0),
\label{SLgen}
\eea
where $M$ is an integer, and $\phi_s^\gamma(0)$ and $\phi^\gamma_B(0)$ are system and environment operators, respectively.
The reduced density matrix at a time $t$ can then be written as 
\bea
\rho_s(t)=\Ttr_B\{{\cal U}_I(t)\rho_\ttot(0){\cal U}_I^\dagger(t)\}=\sum_{\gamma=1}^M  \rho_s^\gamma(t),
\label{rhos_gen}
\eea
where we have defined $\rho_s^\gamma(t)=\Ttr_B\{{\cal U}_I(t)\phi_s^\gamma(0)\otimes\phi^\gamma_B(0) {\cal U}_I^\dagger(t)\}$, with ${\cal U}_I (t,0)=e^{iH_B t}e^{-iH_{\ttot}t}$ the evolution operator in the interaction picture. 
Let us now consider a Bargmann coherent state basis for the environment \cite{bargmann1961}, so that each state is given as a tensorial product of the states of all the oscillators $|z_0 \rangle=|z_{0,1}\rangle| z_{0,2}\rangle...| z_{0,\lambda}\rangle...$, and define 
\bea
\phi^\gamma_B={\mathcal M}_{00'}[f^{\gamma}(z^*_0,z'_0)|z_0\rangle\langle z'_0|],
\label{envcoefSLD}
\eea
where we have considered the closure relation of the Bargmann basis ${\mathcal M}_0[|z_0\rangle\langle z_0|]=1$, and the definitions  ${\mathcal M}_0[\cdots]=\int d\mu(z_0)\cdots$, and ${\mathcal M}_{00'}[\cdots]=\int d\mu(z_0)\int d\mu(z'_0)\cdots$, with the Gaussian measure $d\mu(z_0)=\prod_{\lambda} \frac {d^2 z_{0,\lambda} }{\pi}e^{-|z_{0,\lambda}|^2}$. Also, in Eq. (\ref{envcoefSLD}), we have introduced the functions $f^{\gamma}(z^*_0,z'_0)=\langle z_0|\phi_B^\gamma|z'_0\rangle$. 
Hence, we can write the reduced density operator as in Eq. (\ref{rhos_gen}), where now 
\bea
\rho^\gamma_s(t)={\mathcal M}_{00'}[\rho^\gamma_s(z_0,z'_0|t)],
\label{rhosgamma}
\eea
with $\rho^\gamma_s(z_0,z'_0|t)=$ 
$f^\gamma(z^*_0,z'_0){\mathcal M}_{0''}[G(z_0^{''*},z_0|t,0)\phi^\gamma_s(0)$\\$\times G^{\dagger}(z'^{*}_0,z''_0|t,0)]$, written in terms of the reduced propagators $G(z''^{*}_0,z_0|t,0)=\langle z''_0|{{\cal U}}_I (t,0)|z_0\rangle$. Such quantities, introduced in \cite{strunz2001} and further generalized in \cite{alonso2005,devega2006}, are the matrix elements of the total evolution operator in the Bargmann coherent basis of the environment. Hence, they are defined in the Hilbert space of the system, representing transitions between the environmental state $|z_0\rangle$ to $|z''_0\rangle$. 
The evolution equation for each coefficient, $\rho^\gamma_s(z_0,z'_0|t)$ can be written as (see Supplementary Material (SM) for details),
\begin{align}
&\frac{d\rho_s^\gamma(z_0,z'_0 |t)}{dt}=-i[H_S ,\rho_s^\gamma(z_0,z'_0 |t)]+\cr
&z_{0,t}[\rho_s^\gamma(z_0,z'_0 |t) ,L^{\dagger}]+z'^*_{0,t}[L,\rho^\gamma(z_0,z'_0 |t)]+\nonumber\\
&\bigg(\int_0^t d\tau\alpha(t-\tau)[V_{\tau-t}L\rho_s^\gamma(z_0,z'_0 |t) ,L^{\dagger}]+\Hhc\bigg),
\label{ic6evol0}
\end{align}
up to third order in the perturbative parameter, with an initial condition given by 
\begin{eqnarray}
\rho^\gamma(z_0,z'_0 |0)&=&f^\gamma(z_0,z'_0){\mathcal M}_{0''}[e^{z''^*z_0}e^{{z'}_0^{*}z''}] \phi^\gamma_s(0)\cr
&=&f^\gamma(z_0,z'_0)e^{{z'}_0^{*}z_0}\phi^\gamma_s(0),
\end{eqnarray}
where we have used $G(z''^*_0,z_0|0,0)=\langle z''_0|z_0\rangle=e^{z''^*_0z_0}$.
In Eq. (\ref{ic6evol0}) we have also defined $V_tA=e^{iH_St}Ae^{-iH_St}$ for any system operator $A$, $z_{0,t}=i\sum_\lambda g_\lambda z_{0,\lambda}e^{-i \omega_\lambda t}$, where $z_{0,\lambda}$ are the complex eigenvalues of the annihilation operator, $a_\lambda |z_{0,\lambda}\rangle=z_{0,\lambda}|z_{0,\lambda}\rangle$, and $\alpha(t,\tau)=
\sum_\lambda g_\lambda^2 e^{-i\omega_\lambda (t-\tau)}$, which is the environment correlation function.

In order to proceed further and derive the evolution of Eq. (\ref{rhosgamma}), it is necessary to compute the average ${\mathcal M}_{00'}[\cdots]$ of Eq. (\ref{ic6evol0}). The resulting equation for $\rho_s^\gamma(t)$ can be written as (see SM)
\begin{align}
&\frac{d\rho_s^\gamma(t)}{dt}=-i[H_S ,\rho_s^\gamma(t)]+\cr
&A^{\gamma}(t)[V_t\phi^\gamma_s(0),L^\dagger]+\hat{A}^{\gamma}(t)[L,V_t\phi^\gamma_s(0)]+\cr
&\bigg(\int_0^t d\tau B^\gamma(t,\tau)[[V_{\tau-t}L,\rho_s^{\gamma}(t)],L^{\dagger}] +\cr
&\int_0^t d\tau C^\gamma(t,\tau)[L,[\rho_s^{\gamma}(t),V_{\tau-t}L^\dagger]]+ \cr
&\int_0^t d\tau\alpha(t-\tau)[V_{\tau-t}L\rho^\gamma_s(t) ,L^{\dagger}]+\Hhc\bigg),
\label{general_evolSLD}
\end{align}
where we have defined a series of correlation functions $A^{\gamma}(t)=
i\sum_{q}g_qe^{-i\omega_q t}{\mathcal V}_{Aq}^{\gamma}$, $\hat{A}^{\gamma}(t)=
-i\sum_{q}g_qe^{i\omega_q t}\hat{{\mathcal V}}_{Aq}^{\gamma}$, $B^\gamma(t,\tau)=\sum_{qq'}g_qg_{q'}e^{-i\omega_q t}e^{i\omega_{q'} \tau}{\mathcal V}_{Bqq'}^\gamma$, and $C^\gamma(t,\tau)=-\sum_{qq'}g_qg_{q'}e^{-i\omega_q t}e^{-i\omega_{q'} \tau}{\mathcal V}_{Cqq'}^\gamma$, with 
%
\bea
{\mathcal V}_{Aq}^\gamma&=&{\mathcal M}_{00'}[z_{0q}f^{\gamma}(z^*_0,z'_0)e^{z'^{*}_0z_0}],\cr 
\hat{{\mathcal V}}_{Aq}^{\gamma}&=&{\mathcal M}_{00'}[z'^*_{0q}f^{\gamma}(z^*_0,z'_0)e^{z'^{*}_0z_0}],\cr 
{\mathcal V}_{Bqq'}^\gamma&=&{\mathcal M}_{00'}[z_{0q}z'^*_{0q'}f^{\gamma}(z^*_0,z'_0)e^{z'^{*}_0z_0}],\cr 
{\mathcal V}_{Cqq'}^\gamma&=&{\mathcal M}_{00'}[z_{0q}z'_{0q'}f^{\gamma}(z^*_0,z'_0)e^{z'^{*}_0z_0}].
\label{uves}
\eea
For reasons that will become clear later, we shall expand $B^\gamma(t,\tau)=B^{\gamma,\textmd{neq}}(t,\tau)+B^{\gamma,\textmd{eq}}(t-\tau)$, where $B^{\gamma,\textmd{neq}}(t,\tau)=\sum_{qq'}(1-\delta_{qq'})g_qg_{q'}e^{-i\omega_q t}e^{i\omega_{q'} \tau}{\mathcal V}_{Bqq'}^\gamma$ and $B^{\gamma,\textmd{eq}}(t-\tau)=\sum_{q}g^2_qe^{-i\omega_q (t-\tau)}{\mathcal V}_{Bqq}^\gamma$.
Note that Eq. (\ref{general_evolSLD}) involves certain correlation functions that depend on $t$ and $\tau$, but not on their difference $t-\tau$. Hence, as we will see below, these functions will vanish for environments at equilibrium. 
The form of the functions (\ref{uves}) for a general environment operator, $\phi^\gamma_B=\sum_{\bf np}c^\gamma_{\bf np}|{\bf n}\rangle\langle {\bf p}|$, where $|{\bf n}\rangle$ is the Fock basis and $H_B$ is diagonal is discussed in detail in the SM. This allows to determine for which particular initial states the coefficients (\ref{uves}), and therefore the correlation functions, are non-vanishing.

\textit{Different types of initial state---}
Let us consider first \textit{special linear states (SL)}, which, following the definition in \cite{shabani2009}, are states of the general form (\ref{SLgen}) where we can re-write $\phi^\gamma_B(0)=\rho^\gamma_B(0)$, and $\phi_s^\gamma(0)=\rho_s^\gamma(0)$, since they correspond to density matrices. Thus, the reduced density operator of the system at $t=0$ is $\rho_s(0)=\sum^M_{\gamma=1}\phi_s^\gamma(0)$. Also,  the dynamics given by (\ref{general_evolSLD}) can be represented by a Hermitian map $\Lambda^\gamma_{H}(t)$ such that 
\bea
\rho_s(t)=\sum_\gamma \Lambda^\gamma_{H}(t)\rho^\gamma_s(0),
\eea
with $\Lambda^\gamma_{H}(t)=\unit_S$. We may additionally consider two different situations, (a) when some $\rho^\gamma_B(0)$ are not diagonal in the basis of $H_B$: for those terms we find that $\Ttr_B[B\rho^\gamma_B(0)]\neq 0$, since $B$ is an operator that in general represents transitions between states in the environment energy basis; 
or (b) when all $\rho^\gamma_B(0)$ are diagonal in the basis of $H_B$, and thus $\Ttr_B[B\rho^\gamma_B(0)]= 0$. 
This corresponds to a situation where each $\rho^\gamma_B(0)$ represents the density operator of an environment \textit{in equilibrium}. Since each $\rho^\gamma_B(0)$ can be expressed in terms of a $f^{\gamma}(z^*_0,z'_0)$ that contains the same powers of $z_0$ and ${z'}_0^{*}$, the corresponding $\rho_s^\gamma(t)$ has an evolution equation simpler than Eq. (\ref{general_evolSLD}):
\bea
&&\frac{d\rho_s^\gamma(t)}{dt}=-i[H_S ,\rho_s^\gamma(t)]\cr
&+&\bigg(\int_0^t d\tau B^{\gamma,\textmd{eq}}(t-\tau)[[V_{\tau-t}L,\rho_s^{\gamma}(t)],L^{\dagger}]\cr
&+&\int_0^t d\tau\alpha(t-\tau)[V_{\tau-t}L\rho^\gamma_s(t) ,L^{\dagger}]+\Hhc\bigg),
\label{general_evolSLVD}
\eea 
where now all correlations depend on time differences $t-\tau$ as is typical of environments in equilibrium.

Let us now consider \textit{non-special linear states (NSL)}, which can be written as Eq. (\ref{SLgen}), but with $\Ttr_s\{\phi^\gamma_s(0)\}=\Ttr_B\{\phi^\gamma_B(0)\}= 0$ for some $\gamma$. As such, and contrary to SL states, NSL states do not fulfill the property that the reduced density operator of the system at $t=0$ is a sum of all $\phi_s^\gamma(0)$. Indeed, the initial reduced density matrix is a limited sum $\rho_s(0)=\sum^N_{\gamma=1}\phi_s^\gamma(0)$, where $N< M$, while $\rho_s(t)=\sum^M_{\gamma=1}\rho^\gamma_s(t)$. 
Hence, in this case we can not consider a Hermitian map $\Lambda_\gamma(t)$ such that $\rho^\gamma_s(t)=\Lambda_\gamma(t)\rho^\gamma_s(0)$ for $\gamma=1,\cdots,M$ as before, as that would require having $\Lambda_\gamma(0)=0$ for all the $\gamma>N$, which would contradict the property of linear maps, $\Lambda_\gamma(0)=\unit_S$. 

For initially decorrelated states $\rho_\ttot(0)=\rho_s\otimes\rho_B^\tth$, with $\rho_B^\tth$ a thermal equilibrium state, it is known that the Markov and secular approximations on a second-order weak-coupling master equation lead to a Lindblad form \cite{breuerbook,devega2015c}. When considering these approximations in the more general second-order Eq. (\ref{general_evolSLD}), we find that
\bea
&&\frac{d\rho_s^\gamma(t)}{dt}=\sum_\omega \gamma_A(\omega)[V_t\phi^\gamma_s(0),L^\dagger(\omega)]\cr
&+&\sum_\omega \hat{\gamma}_A(\omega)[L(\omega),V_t\phi^\gamma_s(0)]\cr
&+&\bigg(\sum_\omega \gamma_B(\omega,\omega')[[L(\omega),\rho_s^{\gamma}(t)],L^{\dagger}(\omega')]\cr
&+&\sum_\omega \gamma_C(\omega,\omega')[L(\omega'),[\rho_s^{\gamma}(t),L^\dagger(\omega)]]\cr
&+&\sum_\omega \gamma(\omega)[L(\omega)\rho^\gamma_s(t),L^{\dagger}(\omega)]+\Hhc\bigg).
\label{general_evolSLD_inter2}
\eea 
Here we make use of the spectral decomposition of the coupling operators $L(\omega)=\sum_{\epsilon-\epsilon'=\omega}\Pi(\epsilon)L\Pi(\epsilon')$, with $\Pi(\epsilon)$ representing a projection onto the eigenspace belonging to the eigenvalue $\epsilon$ of $H_S$, which is assumed to have a discrete spectrum. These coefficients and the derivation of Eq. (\ref{general_evolSLD_inter2}) can be found in the SM. The relevant observation here is that, in general, only when $\gamma_A(\omega)=\gamma_C(\omega,\omega')=\gamma_B^{\textmd{neq}}(\omega,\omega')=0$, with $\gamma_B(\omega,\omega')=\gamma_B^{\textmd{eq}}(\omega)+\gamma_B^{\textmd{neq}}(\omega,\omega')$, Eq.  (\ref{general_evolSLD_inter2}) acquires a Lindblad form 
\begin{align}
\frac{d\rho_s^\gamma(t)}{dt}&=\bigg(\sum_\omega \gamma_B^{\textmd{eq}}(\omega)[[L(\omega),\rho_s^{\gamma}(t)],L^{\dagger}(\omega)]\cr
&+\sum_\omega \gamma(\omega)[L(\omega)\rho^\gamma_s(t),L^{\dagger}(\omega)]+\Hhc\bigg).
\label{Linb}
\end{align}
Of the three cases discussed above, this one corresponds to that of an environment in an equilibrium state. 
In the following, we illustrate our formalism for two different examples. 
\begin{figure}[ht]
\centerline{\includegraphics[width=0.45\textwidth]{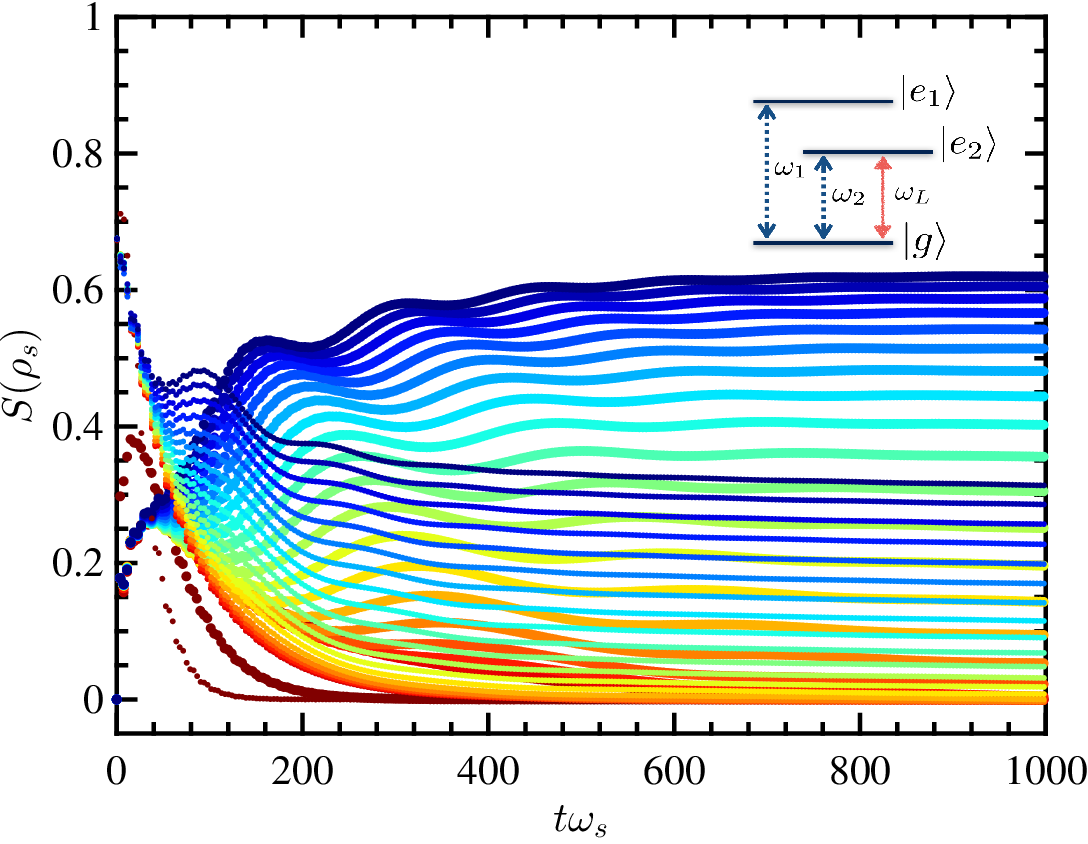}}
\caption{Evolution of the von-Neumann entropy for pure states of the form $\rho_\ttot^{\Nsl}(0)$ (thick lines; finite initial entropy), and $\rho_\ttot^{\Sl}(0)$, (thin lines; zero initial entropy), for increasing values of the laser intensity ranging from $\epsilon=0$ (red) to $0.02$ (violet) in steps of $0.001$, with $\sigma=\sqrt{0.1}$ in Eq. (\ref{distrib}), ${\mathcal A}=\sqrt{0.5}$, ${\mathcal B}=\sqrt{0.25}$, and ${\mathcal C}=\sqrt{0.25}$. The frequency transitions are $\omega_1=1$, and $\omega_2=0.5$, and the parameters for the spectral density $J(\omega)$ are $\alpha=0.005$, $\omega_c=5$. The frequency window is chosen up to $\omega_\mmax=100$. All frequencies are in units of $\omega_s=1$. \label{threelevel}}
\end{figure}

\textit{Example 1---} Let us consider a three-level atom in a $V$-configuration, shown in Fig. \ref{threelevel}, that is coupled to the electromagnetic field, with the transition $|g\rangle\leftrightarrow |e_2\rangle$ driven by a laser with frequency $\omega_L$ and coupling $\epsilon_L$. In the laser rotating frame, the Hamiltonian is given by Eq. (\ref{chapdos1}), with $H_S=\sum_j\Delta_j\sigma_j^+\sigma_j+\epsilon_L(\sigma_2+\sigma_2^+)$, $H_B=\sum_k\Delta_k a_k^\dagger a_k$, and $L=\sigma_1+\sigma_2$. Here, we have defined $\Delta_j=\omega_j-\omega_L$, and $\Delta_k=\omega_k-\omega_L$, with $\omega_j$ and $\sigma_j=|g\rangle\langle e_j|$ corresponding respectively to the transition frequencies and transition operators from each of the two excited states $|e_j\rangle$ to the single ground state $|g\rangle$. The environment is characterized by a linear dispersion relation $\omega_k=ck$ (we set $c=1$), and a sub-ohmic spectral density $J(\omega)=\alpha\omega^{1/2}e^{-\omega/\omega_c}$.

We first consider an initially entangled state with the environment, which is a NSL of the form $\rho_\ttot^{\Nsl}(0)=|\Psi_0\rangle\langle\Psi_0|$, with $|\Psi_0\rangle={\mathcal A}\sigma_1^+|g\rangle|0\rangle+{\mathcal B}\sigma_2^+|g\rangle|0\rangle+{\mathcal C}\sum_k {\mathcal G}_k|g\rangle|1_k\rangle$, where $|1_k\rangle$ is the state with zero excitations in all modes and one excitation in the mode with frequency $\omega_k$, and 
\bea
{\mathcal G}_k=\sqrt{\frac{e^{-(k-k_0)^2/2\sigma^2}}{\sigma 2\pi}},
\label{distrib}
\eea 
with $\sigma$ the Gaussian width. Also, the normalization condition requires that $|{\mathcal A}|^2+|{\mathcal B}|^2+ \sum_k|{\mathcal G}_k|^2|{\mathcal C}|^2=1$. This state, expressed in a coherent state basis, is
\bea
|\Psi_0\rangle=\int d\mu(z_0)|\psi(z_0)\rangle|z_0\rangle,
\label{gen_entang}
\eea
with $|\psi(z_0)\rangle={\mathcal A}|e_1\rangle+{\mathcal B}|e_2\rangle+ {\mathcal C}\sum_k {\mathcal G}_k z^*_{0k}|g\rangle$. 
The resulting initial density matrix can be written as Eq. (\ref{SLgen}), with $f^{0}(z^*_0,z'_0)=\sum_{kk'}{\mathcal G}_k {\mathcal G}^*_{k'}z^*_{0k}z'_{0k'}$, $f^{1}(z^*_0,z'_0)=1$, 
$f^{2}(z^*_0,z'_0)=\sum_{k}{\mathcal G}^*_k z'_{0k}$, and $f^{3}(z^*_0,z'_0)=\sum_{k}{\mathcal G}_k z^*_{0k}$.
Then, we consider a SL $\rho^{\Sl}_\ttot(0)=|\psi_s\rangle\langle\psi_s|\otimes |\psi_b\rangle\langle\psi_b|$, where $|\psi_s\rangle={\mathcal A}\sigma_1^+|g\rangle+{\mathcal B}\sigma_2^+|g\rangle+{\mathcal C}|g\rangle$, and $|\psi_b\rangle=(1/\sqrt{2})(|0\rangle+\sum_k {\mathcal G}_k|1_k\rangle)$. 
The evolution equation for the reduced density matrix $\rho_s(t)$ is also given by an equation of the general form (\ref{general_evolSLD}) (see SM for details). Fig. \ref{threelevel} shows the evolution of the entanglement between the $V-$atom and its environment, as measured by the von-Neumann entropy, for these two types of initial condition. Different curves correspond to different laser intensities $\epsilon$ in a rainbow scale, where the larger the intensity the more blue the curves. As can be observed, the two initial conditions lead to dramatically different dynamics and steady states: while $\rho_\ttot^{\Sl}(0)$ produces an entanglement growth, $\rho_\ttot^{\Nsl}(0)$ shows an entanglement decay from the initial value.

\begin{figure}[ht]
\centerline{\includegraphics[width=0.45\textwidth]{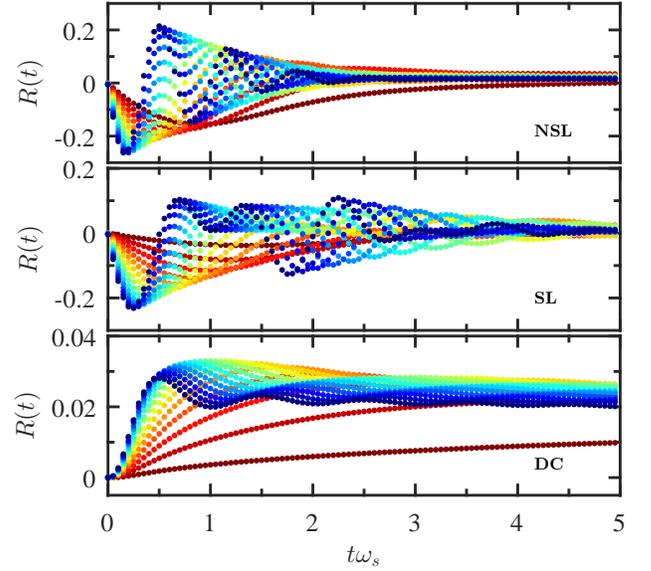}}
\caption{Evolution of the emission rate $R(t)$ for an initial state $\rho^{\Nsl}_\ttot(0)$ (top panel), and a SL state, $\rho^\Sl_\ttot(0)$ (middle panel). Also, ${\mathcal G}^j_k$ depends on $\sigma_q=1$, and $\sigma_2=\sqrt{0.5}$. The coupling parameter is $\alpha=0.05$ for both NSL and SL cases. The coefficients are ${\mathcal A}_1=\sqrt{0.5}$ and ${\mathcal B}_1=\sqrt{0.5}$, ${\mathcal A}_2=\sqrt{0.8}$ and ${\mathcal B}_2=\sqrt{0.20})$. The bottom panel corresponds to the decorrelated (DC) initial state $\rho_\ttot^{\text{DC}}(0)=|\psi_0\rangle\langle\psi_0|\otimes|0\rangle\langle 0|$ with $|\psi_0\rangle={\mathcal A}|g\rangle+{\mathcal B}|e\rangle$, and ${\mathcal A}={\mathcal B}=0.5$. The different curves in all panels correspond to different values of the atomic frequency $\omega_1$ ranging from $0.1$ (red) to $3.1$ (violet) in steps of $0.2$, where $\omega_s$ is our unit, and we consider a sub-ohmic spectral density with $\alpha=0.005$.\label{secondset1}}
\end{figure}

\textit{Example 2---} We now analyze the dynamics of a two-level system with states $|g\rangle$ and $|e\rangle$ coupled to a bosonic field, considering a total Hamiltonian (\ref{chapdos1}) with $H_S=\omega_1\sigma^+\sigma$, $H_B=\sum_k\omega_k a_k^\dagger a_k$, and $L=\sigma+\sigma^+$, with $\sigma=|g\rangle\langle e|$.  We first consider an initially entangled mixed state of the form $\rho^{\Nsl}_\ttot(0)=\sum_{j=1}^2\rho_\ttot^j(0)$, with $\rho_\ttot^j(0)=|\Psi^j_0 \rangle\langle\Psi^j_0|$, and $|\Psi^j_0\rangle={\mathcal A}_j\sum_k {\mathcal G}^j_k |g\rangle |1_k\rangle+{\mathcal B}^j |e\rangle  |0\rangle$. Here, ${\mathcal G}^j_k$ is given by Eq. (\ref{distrib}) with a $j-$dependent Gaussian amplitude $\sigma_j$. 
Expressed in a coherent state basis, each $j$ is given by the general expression (\ref{gen_entang}), with coefficients $|\psi^j(z_0)\rangle= {\mathcal A}_j\sum_k {\mathcal G}_k z^*_{0k}|g\rangle+{\mathcal B}_j|e\rangle$. Note that this state can also be written in the general form (\ref{SLgen}), $
\rho^{\Nsl}_\ttot(0)=\sum_{j=1}^2\sum^3_{\gamma=0} \phi_s^{j\gamma}(0)\otimes\phi^{j\gamma}_B(0)$,
with  $\phi_s^{j0}(0)=|{\mathcal A}_j|^2|g\rangle\langle g|$, $\phi_s^{j1}(0)=|{\mathcal B}_j|^2|e\rangle\langle e|$, $\phi_s^{j2}(0)={\mathcal B}^*_j{\mathcal A}_j|g\rangle\langle e|$, and $\phi_s^{j3}(0)={\mathcal B}_j{\mathcal A}^*_j|e\rangle\langle g|$; and $\phi^{j\gamma}_B(0)$ given by Eq. (\ref{envcoefSLD}) with the same coefficients $f^{j\gamma}$ as in the first example. Indeed, this state is NSL, since $\Ttr_B\{\phi^{j\gamma}_B(0)\}=0$ for $\gamma=2,3$ in both $j$.
The normalization condition is $\sum_j(|{\mathcal B}_j|^2+ \sum_k|{\mathcal G}^j_k|^2|{\mathcal A}_j|^2)=1$. 
For the same system, we now consider an initial state of the form $\rho^\Sl_\ttot(0)=\sum_{j=1}^2 \rho^{j}_s \otimes \rho^{j}_B$ where $\rho^{j}_B=|\psi_b^j\rangle\langle\psi_b^j|$, and $\rho_s^j=|\psi_0^j\rangle\langle\psi_0^j|$, with $|\psi^j_0\rangle={\mathcal A}^j|g\rangle+{\mathcal B}^j|e\rangle$, and $|\psi^j_b\rangle=(1/\sqrt{2})(|0\rangle+\sum_k {\mathcal G}^j_k|1_k\rangle)$. Let us consider the evolution of the emission rate $R(t)=-d\langle\sigma_z(t)\rangle/dt$, with $\sigma_z=2\sigma^+\sigma+1$. In the Markov limit and for a single atom 
coupled to an environment in equilibrium, this quantity has a positive value at $t=0$ and thereafter decays until the atomic population reaches its steady state. Non-Markovian effects lead to an initial increase of $R(t)$, due to the increase of the atomic dissipation rate from zero to a finite value (see bottom panel of Fig. \ref{secondset1}).  
However, as observed in Fig. \ref{secondset1}, the presence of initial correlations (top and middle panels) leads to a rate that initially acquires negative values, which produces an initial gain in populations. A similar increase of populations (and thus negative values of $R(t)$) can also be observed when considering zero correlations but an environment not in equilibrium (see example 3 in SM).

In summary, we have derived a general set of equations that characterize the evolution of an OQS considering an arbitrary total (system and environment) initial state. These evolution equations contain not only the terms that appear in the master equation for initially decorrelated states, but also other terms that depend on new correlation functions. These correlations depend on the coefficients (\ref{uves}), which are non-vanishing only for certain types of initial states. We note that the practicality of the derived equation is based on the fact that these coefficients are computed, for each case, by solving well-known Gaussian integrals. In addition, we  have shown that only initial states of the general form (\ref{SLgen}), with $\Ttr_B\{\phi^\gamma_B(0)\}=1$ (\emph{i.e.} SL) and $\Ttr_B\{B\phi_B^\gamma(0)\}=0$ for all $\gamma$, give rise to a Lindblad equation under the secular and Markov approximations. Finally, we have illustrated the use of our formalism with two different examples: a driven three-level atom in a $V-$configuration, while considering two different types of initial pure states, entangled and non-correlated; and a two-level atom, while considering an initially entangled mixed state and an initially correlated mixed state.   

\textit{Acknowledgments}
The authors are grateful to D. Alonso, M.C. Ba\~nuls, A. Rivas and U. Schollw{\"o}ck for interesting discussions. 
This project was financially supported by the Nanosystems Initiative Munich (NIM) (project No. 862050-2) and partially from the Spanish MINECO through project FIS2013-41352-P and COST Action MP1209.

\bibliography{/Users/ines.vega/Dropbox/ReviewDaniel/Bibtexelesdrop_d1}
\bibliographystyle{prsty}

\end{document}


\title[]{Supplementary material: A weak coupling master equation for arbitrary initial conditions}
\author{Jad C. Halimeh}
\affiliation{Department of Physics and Arnold Sommerfeld Center for Theoretical Physics, Ludwig-Maximilians-Universit{\"a}t M{\"u}nchen, Theresienstr. 37, 80333 Munich, Germany}
\author{In\'es de Vega}
\affiliation{Department of Physics and Arnold Sommerfeld Center for Theoretical Physics, Ludwig-Maximilians-Universit{\"a}t M{\"u}nchen, Theresienstr. 37, 80333 Munich, Germany}

\maketitle

\section{Derivation of Eq. (6)}
Let us first derive Eq. (6) in the main text. To this order, and following a procedure similar to the one in \cite{devega2006}, we consider the evolution equation for the propagator $G(z''^*_0, z_0 |t,0)$,
\begin{eqnarray}
&&\frac{dG(z_i^*, z_{i+1} |t,0)}{dt}=-iH_S G(z_i^*, z_{i+1} |t,0)+L z^*_t  G(z_i^*, z_{i+1} |t,0)\cr
&-&L^\dagger z_{i+1,t}G(z_i^*, z_{i+1} |t,0)\cr
&-& L^\dagger \int^t_0 d\tau \alpha(t-\tau) V_{\tau-t}L G(z_i^*, z_{i+1} |t,0)+{\mathcal O}(g^3).
\label{linearNM2b}
\end{eqnarray}
with $z_{i+1}=z_0$ and $z_i =z''_0$. In this equation, we have defined 
the functions $z_t$ as
\begin{equation}
z_{i,t}=i\sum_\lambda g_\lambda z_{i,\lambda}e^{-i \omega_\lambda t},
\label{chapdos14b}
\end{equation}
where $z_{i,\lambda}$ are the complex numbers such that $a_\lambda |z_{i,\lambda}\rangle=z_{i,\lambda}|z_{i,\lambda}\rangle$, and $\alpha(t,\tau)={\mathcal M}[z_{i,t}z^*_{i,\tau}]=\sum_\lambda g_\lambda^2 e^{-i\omega_\lambda (t-\tau)}$ is the environment correlation function. Considering (\ref{linearNM2b}),  the evolution equation of $\rho_s (z_0,z'_0 |t)$ can be written as 
\begin{eqnarray} 
&&\frac{d\rho^\gamma (z_0,z'_0 |t)}{dt} =-i[H_S ,\rho^\gamma (z_0,z'_0 |t)]-z_{0,t}L^{\dagger}\rho_s (z_0,z'_0 |t) -z'^{*}_{0,t}\rho_s (z_0,z'_0 |t) L \nonumber\\
&+&\int_0^t d\tau \alpha(t-\tau)L^{\dagger}V_{\tau-t}L\rho^\gamma (z_0,z'_0 |t) -\int_0^t d\tau \alpha^* (t-\tau)\rho^\gamma (z_0,z'_0 |t) V_{t-\tau} L^{\dagger}L\nonumber\\
&+&L{\cal M}_{0''} \left[{z''}_{0,t}^{*} G(z''^*_0, z_0 |t,0)| \psi_0 (z^*_0)\rangle\langle\psi_0 (z'_0) |  G^{\dagger}(z'^{*}_0,z''_0|t,0)\right]\cr
&+&{\cal M}_{0''} \left[z''_{0,t} G(z''^*_0, z_0 |t,0)| \psi_0 (z^*_0)\rangle\langle\psi_0 (z'_0) |  G^{\dagger}(z'^{*}_0,z''_0|t,0)\right]L^{\dagger}.
\label{ic2}
\end{eqnarray}
In order to obtain the master equation up to second order in the perturbative parameter, we can use the perturbative expansion in the interaction picture with respect to the system:
\begin{eqnarray}
&&G(z''^*_0,z_0 |t_1,0)=\bigg\{1+\int^{t_1}_{0} d\tau z''^*_{0,\tau}V_{\tau-t_1} L-\int^{t_1}_{0} d\tau z_{0,\tau}V_{\tau-t_1} L^{\dagger}\cr
&-&\int^{t_1}_{0} d\tau\int_0^{\tau} d\tau' \alpha(\tau-\tau')V_{\tau-t_1} L^{\dagger}V_{\tau'-t_1}L+\int^{t_1}_0 d\tau\int_0^{\tau} d\tau' z''^*_{0,\tau}z''^*_{0,\tau'}V_{\tau-t_1} LV_{\tau'-t_1}L\nonumber\\
&-&\int^{t_1}_0 d\tau\int^{\tau}_{0 } d\tau' z''^*_{0,\tau}z_{0,\tau'}V_{\tau-t_1} LV_{\tau'-t_1}L^{\dagger}-\int^{t_1}_{0} d\tau\int_{0}^{\tau} d\tau' z_{0,\tau}z''^*_{0,\tau'}V_{\tau-t_1} L^{\dagger}V_{\tau'-t_1}L\nonumber\\
&-&\int^{t_1}_{0} d\tau\int_{0}^{\tau} d\tau' z_{0,\tau}z_{0,\tau'}V_{\tau-t_1} L^{\dagger}V_{\tau'-t_1}L^{\dagger}\bigg\}G^{(0)}(z''^*_0,z_0 |t_1,0).
\label{perturb2chap3}
\end{eqnarray}
Such an expansion, up to first order in $g$, reads
\begin{eqnarray}
G({z}_0^{''*},z_0 |t,0)=\left(1+\int^t_0 d\tau {z''}^*_{0,\tau} V_{\tau-t}L-\int^t_0 d\tau z_{0\tau}V_{\tau-t}L^\dagger \right)G^{(0)}({z}_0^{''*},z_0|t,0)+{\mathcal O}(g^2);\nonumber\\
G^\dagger ({z'}^*_0, z''_0 |0,t)=G^{\dagger(0)}({z'}_0^*,z''_0|0,t )\left(1-\int^t_0 d\tau z^*_{0,\tau} V_{\tau-t}L+\int^t_0 d\tau {z''}_{0,\tau}V_{\tau-t}L^\dagger \right)+{\mathcal O}(g^2),
\label{firstorderchap3}
\end{eqnarray}
where $G^{(0)}({z}_0^{''*},z_0|t,0)=\exp{(-iH_S t)}\exp{( {z}_0^{''*} z_0)}$, and $G^{\dagger,(0)}({z'}_0^*,z''_0|0,t )=\exp{(iH_S t)}\exp{({z'}^*_0 z''_0)}$, shall be inserted in the average of the noise term $z''_{0,t}$, such that we get
\begin{eqnarray}
&&{\cal M}_{0''} [z''_{0,t} G(z''^*_0, z_0 |t,0)| \psi_0 (z^*_0)\rangle\langle\psi_0 (z'_0) |  G^{\dagger}(z'^{*}_0,z''_0 |t,0)]\cr
&=&\int d\mu(z''_0)\bigg(z''_{0,t} \left(1+\int^t_0 d\tau {z''}^*_{0,\tau} V_{\tau-t}L-\int^t_0 d\tau z_{0\tau}V_{\tau-t}L^\dagger \right)e^{-iH_S t}e^{ {z}_0^{''*} z_0} |\psi_0\rangle\nonumber\\
&&\langle\psi_0|\left(1-\int^t_0 d\tau z^*_{0,\tau} V_{\tau-t}L+\int^t_0 d\tau {z''}_{0,\tau}V_{\tau-t}L^\dagger \right)e^{iH_S t}e^{z'^*_0 z''_0}\bigg)+{\mathcal O}(g^2).
\label{chap3bz1}
\end{eqnarray}
In order to perform the average over $z''_0$ it is necessary to take into account that $z''_{0,t}=i\sum^\nu_{\lambda=1} g_\lambda z''_{0,\lambda} \exp{(-i\omega_\lambda t)}$, so that the average ${\mathcal M}_{0''}$ in fact represents a set of multi-dimensional integrals over each harmonic oscillator. Hence,
\begin{eqnarray}
{\mathcal M}_{0''} \left[z''_{0,t} e^{ {z}_0^{''*} z_0}e^{z'^*_0 z''_0 }\right]=i\sum^\nu_{\lambda=1} g_\lambda e^{-i\omega_\lambda t} \left(\int d\mu(z''_{0,1})e^{ {z''}_{0,1}^* z_{0,1}}e^{z'^*_{0,1} z''_{0,1} } \cdots \right.\nonumber\\
\left.\int d\mu(z''_{0,\lambda} )z''_{0,\lambda}e^{ {z''}_{0,\lambda}^* z_{0,\lambda}}e^{z'^*_{0,\lambda} z''_{0,\lambda}} \cdots \int d\mu(z_{1,\nu}) e^{ {z''}_{0,\nu}^* z_{0,\nu}}e^{z'^*_{0,\nu} z''_{0,\nu}}\right).
\end{eqnarray}
Taking this into account, and the following general identities for Gaussian integrals:
\begin{eqnarray}
&&\int d\mu(z_{1,\lambda})z_{1,\lambda }e^{ z_{1,\lambda}^* z_{0,\lambda}}e^{z^*_{0,\lambda} z_{1,\lambda}}=z_{0,\lambda}e^{ z_{0,\lambda}^* z_{0,\lambda}},\nonumber\\
&&\int d\mu(z^*_{1,\lambda})z^*_{1,\lambda }e^{ z_{1,\lambda}^* z_{0,\lambda}}e^{z^*_{0,\lambda} z_{1,\lambda}}=z^*_{0,\lambda}e^{ z_{0,\lambda}^* z_{0,\lambda}},\nonumber\\
&&\int d\mu(z^*_{1,\lambda})z^*_{1,\lambda }z_{1,\lambda }e^{ z_{1,\lambda}^* z_{0,\lambda}}e^{z^*_{0,\lambda} z_{1,\lambda}}=(1+z^*_{0,\lambda} z_{0,\lambda})e^{ z_{0,\lambda}^* z_{0,\lambda}},
\end{eqnarray}
we get for (\ref{chap3bz1}),
\begin{eqnarray}
{\mathcal M}_1 \left[z''_{0,t} G({z}_0^{''*},z_0 |t,0)|\psi_0 \rangle\langle\psi_0 | G^{\dagger}(z^{*}_0,z''_0|t,0)\right]
=z_{0,t}\rho^{(1)}_s (z^*_0 z_0 |t)+\int^t_0 d\tau \alpha(t-\tau) V_{\tau-t}L\rho^{(0)}_s (z^*_0,z_0 |t), 
\label{chap3bz1b}
\end{eqnarray}
up to second order in the perturbative parameter $g$. The former expression is written in terms of 
\begin{eqnarray}
\rho^{(1)\gamma} (z_0,z'_0 |t)&=&\rho^{(0)}_s (z_0,z'_0 |t)+\int^t_0 d\tau z'^*_{0,\tau}[V_{\tau-t}L,\rho^{(0)\gamma} (z_0,z'_0 |t)]\cr
&+&\int^t_0 d\tau z_{0,\tau} [\rho^{(0)\gamma} (z_0,z'_0 |t),V_{\tau-t}L^\dagger],
\label{ic4}
\end{eqnarray}
which is the first-order expansion of $\rho^\gamma(z_0,z'_0 |t) $.  Note that the last expression is in terms of
\begin{equation}
\rho^{(0)\gamma} (z_0,z'_0 |t)=e^{-iH_S t}|\psi_0 (z_0)\rangle\langle\psi_0 (z'_0)|e^{iH_S t} e^{ z'^*_{0} z_{0}}.
\end{equation}
This quantity appears in a term that is already of first order in $g$, which means that we can consistently replace $\rho^{(1)\gamma} (z_0,z'_0 |t)\approx\rho^\gamma(z_0,z'_0 |t)$. Hence, up to second order, we find 
\begin{align}\nonumber
&{\cal M}_{0''} \left[z''_{0,t} G(z''^*_0, z_0 |t,0)| \psi_0 (z^*_0)\rangle\langle\psi_0 (z'_0) |  G^{\dagger}(z'^{*}_0,z''_0 |t,0)\right]\\
&=z_{0,t}\rho_s (z_0,z'_0 |t)+\int_0^t d\tau \alpha(t-\tau)V_{\tau-t}L \rho_s (z_0,z'_0 |t).
\label{ic3}
\end{align}
In the same way, and once again up to second order in $g$, we have 
\begin{align}\nonumber
&{\cal M}_{0''}\left[{z''}_{0,t}^{*}G(z_0^*,z_0 |t,0)| \psi_0 (z^*_0)\rangle\langle\psi_0 (z'_0) |  G^{\dagger}(z'^{*}_0,z''_0|t,0) \right]\\
&=z'^*_{0,t}\rho_s (z_0,z'^{*}_0|t,0)+\rho_s (z_0,z'_0 |t) \int_0^t d\tau \alpha^*(t-\tau)V_{\tau-t}L^{\dagger}.
\label{ic5}
\end{align}
Inserting (\ref{ic3}) and (\ref{ic5}) in (\ref{ic2}), we get the evolution for $\rho_s (z_0,z'^{*}_0)$ up to second order, as given by Eq. (6) in the manuscript.

\section{Derivation of Eq. (8)}

Let us now derive the evolution equation (8) of the manuscript, by considering the average ${\mathcal M}_{00'}[\cdots]$ of Eq. (6). To this order, we need to solve the terms ${\mathcal M}_{00'}[z_{0,t}\rho^\gamma(z_0,z'_0 |t)]$ and ${\mathcal M}_{00'}[z'^*_{0,t}\rho^\gamma(z_0,z'_0 |t)]$ in Eq. (6). Since the quantities $z_{i,t}$ are already of first order in the weak coupling parameter $g$, we can replace ${\mathcal M}_{00'}[z_{0,t}\rho^\gamma_s (z_0,z'^{*}_0|t)]={\mathcal M}_{00'}[z_{0,t}\rho^{(1)\gamma}_s (z_0,z'^{*}_0|t)]$, where $\rho^{(1)\gamma}_s (z_0,z'^{*}_0|t)$ is the expansion of $\rho^\gamma_s (z_0,z'^{*}_0|t)$ up to first order:
\begin{eqnarray}
&&\rho^{(1)\gamma}_s (z_0,z'_0 |t)=\rho^{(0)\gamma}_s(z_0,z'_0 |t)\cr
&+&\int^t_0 d\tau z'^*_{0,\tau}[V_{\tau-t}L,\rho^{(0)\gamma}_s(z_0,z'_0 |t)]+\int^t_0 d\tau z_{0,\tau} [\rho^{(0)\gamma}_s(z_0,z'_0 |t),V_{\tau-t}L^\dagger],
\label{ic4SLD}
\end{eqnarray}
and where the zero-order terms are
\bea
\rho^{(0)\gamma}_s(z_0,z'_0 |t)=V_t\rho^\gamma_s(z_0,z'_0|0),
\label{rho0}
\eea
with $\rho^\gamma_s(z_0,z'_0|0)=f^{\gamma}(z^*_0,z'_0)V_t\phi_s^{\gamma}(0)e^{z'^{*}_0z_0}$.
Thus, we find, up to second order in $g$, that
 \bea
&&{\mathcal M}_{00'}[z_{0,t}\rho^{(1)\gamma}_s (z_0,z'^{*}_0|t)]={\mathcal M}_{00'}[z_{0,t}\rho^{(0)\gamma}_s(z_0,z'^{*}_0|t)]\cr
&+&\int_0^t d\tau [V_{\tau-t}L,{\mathcal M}_{00'}[z_{0,t}z'^*_{0,\tau}\rho^{(0)\gamma}_s(z_0,z'^{*}_0|t)]]+\int_0^t d\tau [{\mathcal M}_{00'}[z_{0,t}z_{0,\tau}\rho^{(0)\gamma}_s(z_0,z'^{*}_0|t)],V_{\tau-t}L^\dagger].
\label{average2}
 \eea
As before, we solve the three main terms of the last expression as 
\bea
&&{\mathcal M}_{00'}[z_{0,t}\rho^{(0)\gamma}_s(z_0,z'^{*}_0|t)]=A^{\gamma}(t)V_t\phi_s^{\gamma}(0),\cr
&&{\mathcal M}_{00'}[z_{0,t}z'^*_{0,\tau}\rho^{(0)\gamma}_s(z_0,z'^{*}_0|t)]=B^{\gamma}(t,\tau)V_t\phi_s^{\gamma}(0),\cr
&&{\mathcal M}_{00'}[z_{0,t}z'_{0,\tau}\rho^{(0)\gamma}_s(z_0,z'^{*}_0|t)]=C^{\gamma}(t,\tau)V_t\phi_s^{\gamma}(0),
\eea
up to second order in $g$, and the functions $A$, $B$ and $C$ are defined in the main text. 
The resulting equation for $\rho_s^\gamma(t)$ can be written as in Eq. (8) by just considering the fact that the functions $B$ and $C$ are already of second order in $g$ so that it is possible to replace $\rho_s^{(0)\gamma}(t)=V_t\phi_s^{\gamma}(0)=\rho_s^{\gamma}(t)$ in those terms in full consistency with the second-order weak coupling equation. 

\section{General form for the coefficients in Eq. (9) for an arbitrary $\phi_B^\gamma$ and general properties of the Gaussian integrals}
Let us now specify the functions $f(z_0^*,z'_0)$ appearing in the Bargmann representation of a general environment operator $\phi^\gamma_B=\sum_{\bf np}c^\gamma_{\bf np}|{\bf n}\rangle\langle {\bf p}|$, where $|{\bf n}\rangle$ is the Fock basis in which $H_B$ is diagonal. Then we find that 
\bea
f^{\gamma}(z^*_0,z'_0)=\sum_{\bf np}c^\gamma_{\bf np}\langle z_0|{\bf n}\rangle\langle{\bf p}|z'_0\rangle=\sum_{\bf np}c^\gamma_{\bf np}\prod_{\lambda\lambda'}\frac{(z^*_{0,\lambda'})^{p_{\lambda'}}(z'_{0,\lambda})^{n_\lambda}}{\sqrt{n_\lambda!p_{\lambda'}!}}.
\label{coefficientF}
\eea
Correspondingly, we find that the coefficients of Eq. (9) in the main text, namely 
\bea
{\mathcal V}_{Aq}^\gamma&=&{\mathcal M}_{00'}[z_{0q}f^{\gamma}(z^*_0,z'_0)e^{z'^{*}_0z_0}],\cr 
\hat{{\mathcal V}}_{Aq}^{\gamma}&=&{\mathcal M}_{00'}[z'^*_{0q}f^{\gamma}(z^*_0,z'_0)e^{z'^{*}_0z_0}],\cr 
{\mathcal V}_{Bqq'}^\gamma&=&{\mathcal M}_{00'}[z_{0q}z'^*_{0q'}f^{\gamma}(z^*_0,z'_0)e^{z'^{*}_0z_0}],\cr 
{\mathcal V}_{Cqq'}^\gamma&=&{\mathcal M}_{00'}[z_{0q}z'_{0q'}f^{\gamma}(z^*_0,z'_0)e^{z'^{*}_0z_0}],
\label{uves}
\eea
can be written as
\bea
{\mathcal V}^\gamma_{Aq}&=&\sum_{\bf np}c^\gamma_{\bf np}\sqrt{n_q+1}\delta_{n_q+1,p_q}\prod_{\lambda\neq q}\delta_{n_\lambda p_\lambda},\cr
{\mathcal V}^\gamma_{Bqq'}&=&(1-\delta_{qq'})\sum_{\bf np}c^\gamma_{\bf np}\sqrt{(n_q+1)n_{q'}}\delta_{n_q+1,p_q}\delta_{n_{q'}-1,p_{q'}}\prod_{\lambda\neq qq'}\delta_{n_\lambda p_\lambda}+\delta_{qq'}\sum_{\bf np}\delta_{\bf n,p}C_{\bf nn}n_q,\cr
{\mathcal V}^\gamma_{Cqq'}&=&(1-\delta_{qq'})\sum_{\bf np}c^\gamma_{\bf np}\sqrt{(n_q+1)(n_{q'}+1)}\delta_{n_q+1,p_q}\delta_{n_{q'}+1,p_{q'}}\prod_{\lambda\neq qq'}\delta_{n_\lambda p_\lambda}\cr
&+&\delta_{qq'}\sum_{\bf np}c^\gamma_{\bf np}\sqrt{(n_q+2)(n_q+1)}\delta_{p_q,n_q+2}\prod_{\lambda\neq q}\delta_{n_\lambda p_\lambda}.
\eea
From these expressions, it can be seen that in an evolution up to second order in the weak coupling parameter, we can only capture the effects of diagonal elements of $\phi_B^\gamma(0)$ that correspond to transitions between certain Fock states (see Table \ref{tabla}). Similarly, this also means that we will only be able to capture the entanglement between the system and the environment that involves environment states which are related according to the column in the right-hand side of Table \ref{tabla}. We finally note that in the previous calculation and throughout the paper, we have utilized the following Gaussian integrals: 
\bea
&&\int d\mu(z_{0})(z_{0,\lambda}^*)^{p_\lambda} (z_{0,\lambda})^{n_\lambda}=\delta_{n_\lambda p_\lambda}n_\lambda!,\cr
&&\int d\mu(z_{0})e^{{z'}_{0,\lambda}^{*}z_{0,\lambda}}(z_{0,\lambda}^*)^{p_\lambda}= ({z'}_{0,\lambda}^{*})^{p_\lambda},\cr
&&\int d\mu(z'_{0})e^{{z'}_{0,\lambda}^{*}z_{0,\lambda}}(z'_{0,\lambda})^{n_\lambda}= ({z}_{0,\lambda})^{n_\lambda},\cr
&&\int d\mu(z_{0})e^{{z'}_{0,\lambda}^{*}z_{0,\lambda}}(z_{0,\lambda}^*)^{p_\lambda} (z_{0,\lambda})=p_\lambda ({z'}_{0,\lambda}^{*})^{p_\lambda-n_\lambda},
\eea
where, as earlier defined, $\int d\mu(z_0)=\int d\mu(z_{0,1})\cdots\int d\mu(z_{0,\lambda})\cdots$. 
\begin{table*}[ht]
\caption{Transitions corresponding to the off-diagonal elements of the density matrix that can be captured with an OQS evolution up to second order.}
\centering
\setlength{\tabcolsep}{6pt}
\renewcommand{\arraystretch}{1.} 
\begin{tabular}{p{0.15\linewidth}p{0.40\linewidth}}
\hline
Non-vanishing ${\mathcal V}$  & Transition \\
\hline
${\mathcal V}_{Aq}$   & $|n_1,\cdots,n_q+1,\cdots\rangle\longleftrightarrow |{\bf n}\rangle$\\
${\mathcal V}_{Bqq'}\,\,\,(q\neq q')$& $|n_1,\cdots,n_q+1,\cdots\rangle\longleftrightarrow |n_1,\cdots,n_{q'}-1,\cdots\rangle$\\
${\mathcal V}_{Bqq}$& $|{\bf n}\rangle \longleftrightarrow|{\bf n}\rangle$\\
${\mathcal V}_{Cqq'}\,\,\,(q\neq q')$&$|n_1,\cdots,n_q+1,\cdots \cdots,n_{q'}+1,\cdots \rangle\longleftrightarrow|{\bf n}\rangle$\\
${\mathcal V}_{Cqq}$& $|n_1,\cdots,n_q+2,\cdots\rangle\longleftrightarrow|{\bf n}\rangle$\\
\end{tabular}
\label{tabla}
\end{table*}

\section{Derivation of Eq. (12).}

Let us compute the Markov and secular limits of Eq. (8) in order to analyze for which initial states the resulting equation has a Lindblad form. To this order, and following the standard method that is considered for initially decorrelated conditions \cite{breuerbook,devega2015c}, we first re-write Eq. (8) in the interaction picture with respect to the system: 
\bea
\frac{d\rho_s^\gamma(t)}{dt}&=&
A^{\gamma}(t)[V_t\phi_s^{\gamma}(0),V_tL^\dagger]+\hat{A}^{\gamma}(t)[V_tL,V_t\phi_s^{\gamma}(0)]+\bigg(\int_0^t d\tau B^\gamma(t,\tau)[[V_{\tau}L,\rho_s^{\gamma}(t)],V_tL^{\dagger}]\cr
&+&\int_0^t d\tau C^\gamma(t,\tau)[V_tL,[\rho_s^{\gamma}(t),V_{\tau}L^\dagger]]+\int_0^t d\tau\alpha(t-\tau)[V_{\tau}L\rho^\gamma_s(t),V_tL^{\dagger}]+\Hhc\bigg)
\label{general_evolSLD_inter}
\eea 
and consider the spectral decomposition of the coupling operators $L(\omega)=\sum_{\epsilon-\epsilon'=\omega}\Pi(\epsilon)L\Pi(\epsilon')$, where $\Pi(\epsilon)$ represents a projection onto the eigenspace belonging to the eigenvalue $\epsilon$ of $H_S$, which is assumed to have a discrete spectrum. Note also that $L^+(\omega)=L(-\omega)$. In terms of these quantities, Eq. (\ref{general_evolSLD_inter}) can be re-written in the long-time limit as Eq. (12) in the main text, 
where we have defined the coefficients:
\bea
\gamma_A(\omega)&=&i\sum_{q}g_qe^{-i(\omega_q-\omega)t}{\mathcal V}_{Aq},\cr
\gamma_B(\omega,\omega')&=&(1-\delta_{qq'})\sum_{qq'}g_{q'}g_q{\mathcal V}_{Bqq'}e^{-i(\omega_q-\omega')t}\int_0^\infty d\tau e^{i(\omega_{q'}+\omega)\tau}+\delta_{qq'}\sum_{q}g^2_{q}e^{-i(\omega-\omega')t}\int_0^\infty d\tau e^{i(\omega_{q}-\omega)\tau}{\mathcal V}_{Bqq},\cr
\gamma_C(\omega,\omega')&=&\sum_{qq'}g_{q'}g_qe^{-i(\omega_q-\omega')t}\int_0^\infty d\tau e^{-i(\omega_{q'}-\omega)\tau}{\mathcal V}_{Cqq'},\cr
\gamma(\omega)&=&\sum_{q}g^2_{q}e^{-i(\omega-\omega')t}\int_0^\infty d\tau e^{i(\omega_{q}-\omega)\tau}.
\eea
which can be further simplified as 
\bea
\gamma_A(\omega)&=&i\sum_{q}g_q {\mathcal V}_{Aq}\delta(\omega_q-\omega),\cr
\gamma_B(\omega,\omega')&=&\gamma_B^{\textmd{neq}}(\omega,\omega')+\gamma_B^{\textmd{eq}}(\omega,\omega'),\cr
\gamma_C(\omega,\omega')&=&\sum_{qq'}g_{q'}g_q\delta(\omega_q-\omega')\bigg(\delta(\omega_{q'}-\omega)-iP\frac{1}{\omega_{q'}-\omega}\bigg){\mathcal V}_{Cqq'},\cr
\gamma(\omega)&=&\delta(\omega-\omega')\sum_{q}g^2_{q}\bigg(\delta(\omega_{q}-\omega)+i{\mathcal P}\frac{1}{\omega_{q}-\omega}\bigg),
\eea
where ${\mathcal P}$ denotes the Cauchy principal value, and we have defined
\bea
\gamma_B^{\textmd{neq}}(\omega,\omega')&=&(1-\delta_{qq'})\sum_{qq'}g_{q'}g_q{\mathcal V}_{Bqq'}\delta(\omega_q-\omega')\bigg(\delta(\omega_{q'}+\omega)+i{\mathcal P}\frac{1}{\omega_{q'}+\omega}\bigg),\cr
\gamma_B^{\textmd{eq}}(\omega,\omega')&=&\delta_{qq'}\delta(\omega-\omega')\sum_{q}g^2_{q}{\mathcal V}_{Bqq}\bigg(\delta(\omega_{q}-\omega)+i{\mathcal P}\frac{1}{\omega_{q}-\omega}\bigg)=\gamma_B^{\textmd{eq}}(\omega).
\eea

\section{Numerical examples}
\subsection{Example 1}
We now consider an initially entangled state with the environment, $|\Psi_0\rangle={\mathcal A}\sigma_1^+|g\rangle|0\rangle+{\mathcal B}\sigma_2^+|g\rangle|0\rangle+{\mathcal C}\sum_k {\mathcal G}_k|g\rangle|1_k\rangle$, where $|1_k\rangle$ is the state with zero excitations in all modes and one excitation in the mode with frequency $\omega_k$, and 
\bea
{\mathcal G}_k=\sqrt{\frac{e^{-(k-k_0)^2/2\sigma^2}}{\sigma 2\pi}}.
\label{distrib}
\eea 
This state, expressed in a coherent state basis, is
\bea
|\Psi_0\rangle=\int d\mu(z_0)|\psi(z_0)\rangle|z_0\rangle,
\label{gen_entang}
\eea
with $|\psi(z_0)\rangle={\mathcal A}|e_1\rangle+{\mathcal B}|e_2\rangle+ {\mathcal C}\sum_k {\mathcal G}_k z^*_{0k}|g\rangle$. 
The resulting initial density matrix can be written as Eq. (2) in the main text, with $\Phi_B^\gamma(0)$ given by Eq. (4), with the coefficients 
\bea
f^{0}(z^*_0,z'_0)&=&\sum_{kk'}{\mathcal G}_k {\mathcal G}^*_{k'}z^*_{0k}z'_{0k'},\cr
f^{1}(z^*_0,z'_0)&=&1,\cr
f^{2}(z^*_0,z'_0)&=&\sum_{k}{\mathcal G}^*_k z'_{0k},\cr
f^{3}(z^*_0,z'_0)&=&\sum_{k}{\mathcal G}_k z^*_{0k},
\label{f1s}
\eea
for $\gamma=0,\cdots,3$, and $\phi_s^0(0)=|{\mathcal C}|^2|g\rangle\langle g|$, $\phi_s^1(0)=|{\mathcal A}|^2|e_1\rangle\langle e_1|+|{\mathcal B}|^2|e_2\rangle\langle e_2|+{\mathcal A}{\mathcal B}^*|e_1\rangle\langle e_2|+{\mathcal B}{\mathcal A}^*|e_2\rangle\langle e_1|$, $\phi_s^2(0)={\mathcal C}^*{\mathcal A}|e_1\rangle\langle g|+{\mathcal C}^*{\mathcal B}|e_2\rangle\langle g|$, and $\phi_s^3(0)=\phi_s^{2\dagger}(0)$. With this choice, the only non-vanishing coefficients in Eq. (9) are
\bea
{\mathcal V}_{Aq}^3&=&
\sum_k {\mathcal G}_k\delta_{kq},\cr 
{\mathcal V}_{Bqq'}^0&=&
\sum_{kk'}2{\mathcal G}_k{\mathcal G}^*_{k'}\delta_{q'k'}\delta_{qk},
\label{uvesEx1}
\eea
which, when replaced in the definition of the correlation functions in the main text, lead to
\bea
A^3(t)&=&i\sum_k g_k{\mathcal G}_ke^{-i\omega_kt},\cr
\hat{A}^2(t)&=&-i\sum_k g_k {\mathcal G}^*_ke^{i\omega_kt}=A^{3*}(t),\cr
B^0(t,\tau)&=&\sum_{kk'} 2g_k {\mathcal G}_k g_{k'} {\mathcal G}^*_{k'}e^{-i\omega_{k}t}e^{i\omega_{k'}\tau},\cr
B^1(t,\tau)&=&\sum_{k} g^2_{k}e^{-i\omega_{k}(t-\tau)}.
\eea
Now we need to evolve Eq. (8) in the main text, 
which, component-wise, can be written as
\bea
\frac{d\rho_s^0(t)}{dt}&=&-i[H_S ,\rho_s^0(t)]
+\int_0^t d\tau B^0(t,\tau)[[V_{\tau-t}L,\rho_s^{0}(t)],L^{\dagger}]\cr
&+&\int_0^t d\tau\alpha(t-\tau)[V_{\tau-t}L\rho^0(t) ,L^{\dagger}]+\Hhc,\cr
\frac{d\rho_s^1(t)}{dt}&=&-i[H_S ,\rho_s^1(t)]
+\int_0^t d\tau\alpha(t-\tau)[V_{\tau-t}L\rho_s^1(t) ,L^{\dagger}]+\Hhc,\cr
\frac{d\rho_s^2(t)}{dt}&=&-i[H_S ,\rho_s^2(t)]+A^{2}(t)[L,V_{t}\phi_s^{2}(0)]\cr
&+&\int_0^t d\tau\alpha(t-\tau)[V_{\tau-t}L\rho^2(t) ,L^{\dagger}]+\Hhc,\cr
\frac{d\rho_s^3(t)}{dt}&=&-i[H_S ,\rho_s^3(t)]+A^{3}(t)[V_{t}\phi_s^{3}(0),L^\dagger]\cr
&+&\int_0^t d\tau\alpha(t-\tau)[V_{\tau-t}L\rho^3_s(t) ,L^{\dagger}]+\Hhc
\label{evol_ex1}
\eea

\begin{figure}[ht]
\centerline{\includegraphics[width=0.45\textwidth]{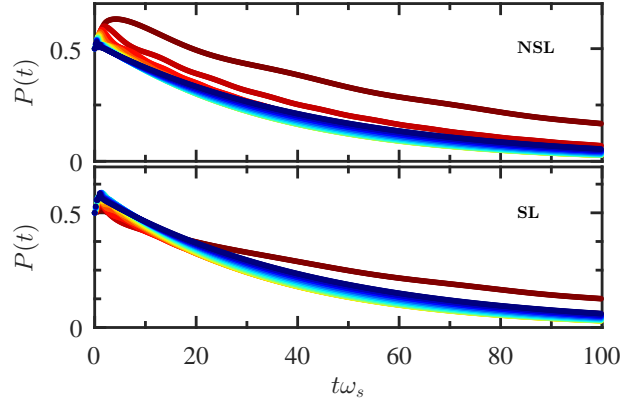}}
\caption{Population results for Example 3: Evolution of the population, $P(t)$, for an initial state $\rho^{\Nsl}_\ttot(0)$ (top panel), and an SL state, $\rho^\Sl_\ttot(0)$ (bottom panel). Also, ${\mathcal G}_k$ depends on $\sigma_q=1$. The coefficients are ${\mathcal A}=\sqrt{0.5}$ and ${\mathcal B}=\sqrt{0.5}$. The different curves correspond to different values of the atomic frequency $\omega_1$ ranging from $0.05$ to $0.85$ in steps of $0.05$. Also, $\omega_s$ is our unit, and we have consider a sub-ohmic spectral density with $\alpha=0.005$. \label{Fig3a}}
\end{figure}
\begin{figure}[ht]
\centerline{\includegraphics[width=0.45\textwidth]{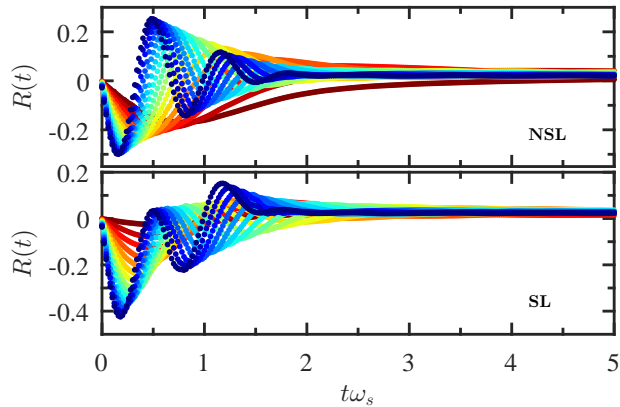}}
\caption{Rate results for Example 3: Evolution of the emission rate $R(t)$ for the same parameters and cases of Fig. \ref{Fig3a}. \label{Fig3b}}
\end{figure}

If we now consider an initial state $\rho^{\text{SL}}_\ttot(0)=\rho_s (0)\otimes \rho_B(0)$, 
where $\rho_B(0)={\mathcal M}_{00'}[\sum_\gamma f^{\gamma}(z_0^*,z'_0)|z_0\rangle\langle z'_0|]=\sum_\gamma\phi_B^{\gamma}(0)$, and $\rho_s(0)=|\psi_0\rangle\langle\psi_0|$, with $|\psi_0\rangle={\mathcal A}|e_1\rangle+{\mathcal B}|e_2\rangle+{\mathcal C}|g\rangle$. We note that in contrast to the previous case, this state is a special linear state, since $\Ttr_B\{\rho_B\}=1$. 
The evolution equation for the reduced density matrix $\rho_s(t)$ is given by an equation of the general form 
\bea
\frac{d\rho_s(t)}{dt}&=&-i[H_S ,\rho_s(t)]\cr
&+&\sum_\gamma A^{\gamma}(t)[V_t\rho_s(0),V_tL^\dagger]+\sum_\gamma \hat{A}^{\gamma}(t)[V_tL,V_t\rho_s(0)]\cr
&+&\bigg(\int_0^t d\tau \sum_\gamma B^\gamma(t,\tau)[[V_{\tau-t}L,\rho_s(t)],L^{\dagger}]\cr
&+&\int_0^t d\tau \sum_\gamma C^\gamma(t,\tau)[L,[\rho_s(t),V_{\tau-t}L^\dagger]]\cr
&+&\int_0^t d\tau\alpha(t-\tau)[V_{\tau-t}L\rho_s(t) ,L^{\dagger}]+\Hhc\bigg).
\label{general_evolSLD2}
\eea 

\subsection{Example 3}
We now analyze the dynamics of a two level system like the one in Example 2 in the main text, \emph{i.e.} considering a total Hamiltonian $H_S + \sum_\lambda \omega_\lambda a_\lambda^\dagger a_\lambda + B^\dagger  L  + B L^\dagger$ with $H_S=\omega_1\sigma^+\sigma$, $H_B=\sum_k\Delta_k a_k^\dagger a_k$, and $L=\sigma+\sigma^+$, with $\sigma=|g\rangle\langle e|$.  We then consider an initially entangled state of the form $\rho^{\Nsl}_\ttot(0)=|\Psi_0 \rangle\langle\Psi_0|$, and $|\Psi_0\rangle={\mathcal A}\sum_k {\mathcal G}_k |g\rangle |1_k\rangle+{\mathcal B} |e\rangle  |0\rangle$, with ${\mathcal G}_k$ given in Eq. (\ref{distrib}). 
For the same system, we also consider an initial separable state of the form $\rho^\Sl_\ttot(0)=\rho_s (0)\otimes \rho_B(0)$ where $\rho_B(0)=|\psi_b\rangle\langle\psi_b|$, and $\rho_s(0)=|\psi_0\rangle\langle\psi_0|$, with $|\psi_0\rangle={\mathcal A}|g\rangle+{\mathcal B}|e\rangle$, and $|\psi_b\rangle=(1/\sqrt{2})(|0\rangle+\sum_k {\mathcal G}_k|1_k\rangle)$. This means that, although the system is initially decorrelated, the environment is not in an equilibrium state in the sense described in the main text, \emph{i.e.} it contains off-diagonal elements in the basis of $H_B$.  As shown in Fig. \ref{Fig3a}, the populations for both initially entangled state $\rho^{\Nsl}_\ttot(0)$, and initially decorrelated (but non-equilibrium) state $\rho^\Sl_\ttot(0)$ present an initial growth just as in Example 2 of the main text. As shown in Fig. \ref{Fig3b}, this corresponds to initial negative values for the decaying rate $R(t)$.

\bibliography{/Users/ines.vega/Dropbox/ReviewDaniel/Bibtexelesdrop_d1}